\documentclass[twocolumn,prb,showpacs]{revtex4}
\usepackage{graphicx,graphics,color,epsfig}
\usepackage{bm}
\usepackage{amsmath}
\usepackage{amssymb}
\usepackage{epstopdf}
\usepackage{dcolumn}

\newcommand{\rhos}{$\rho_{\rm s}(T)$}

\begin{document}


\title{Local suppression of the superfluid density of PuCoGa$_5$ in the Swiss Cheese model}

\author{Tanmoy Das}
\affiliation{Theoretical Division, Los Alamos National Laboratory, Los Alamos, New Mexico, 87545 USA}

\author{Jian-Xin~Zhu}
\affiliation{Theoretical Division, Los Alamos National Laboratory, Los Alamos, New Mexico, 87545 USA}

\author{Matthias J.~Graf}
\affiliation{Theoretical Division, Los Alamos National Laboratory, Los Alamos, New Mexico, 87545 USA}

\date{\today}

\begin{abstract}
We present superfluid density calculations for the unconventional superconductor PuCoGa$_5$ by solving the real-space Bogoliubov-de Gennes equations on a square lattice within the Swiss Cheese model in the presence of strong on-site disorder. We find that despite strong electronic inhomogeneity, one can establish a one-to-one correspondence between the local maps of the density of states, superconducting order parameter, and superfluid density.
In this model, strong on-site  impurity scattering  punches localized holes into the fabric of $d$-wave superconductivity similar to a Swiss cheese. Already a two-dimensional impurity concentration of $n_{\rm imp} = 4\%$ gives rise to a pronounced short-range suppression of the order parameter and a suppression of the superconducting transition temperature $T_c$ by roughly  20\% compared to its pure limit value $T_{c0}$,  whereas the superfluid density $\rho_s$ is reduced drastically by about 70\%.
This result is consistent with available experimental data for aged (400 day-old) and fresh (25 day-old) PuCoGa$_5$ superconducting samples.  In addition, we show that the $T^2-$dependence of the low-$T$ superfluid density, a signature of dirty $d$-wave superconductivity, originates from a combined effect in the density of states of `gap filling' and `gap closing'.
Finally, we demonstate that the Uemuera plot of $T_c$ vs.\ $\rho_s$ deviates sharply from the conventional Abrikosov-Gor'kov theory for radiation-induced defects in PuCoGa$_5$, but follows the same trend of short-coherence-length high-$T_c$ cuprate superconductors.
\end{abstract}

\maketitle


\section{Introduction}

The study of the superfluid density or superfuid stiffness in response to disorder or defects can provide valuable information about the nature of the superconducting order parameter.  The superfluid density $\rho_s$ is proportional to the inverse-square of the penetration depth $\lambda$, a characteristic length of a bulk superconductor determining the penetration of a magnetic field inside it. The low-temperature dependence of $\lambda(T)$ is being studied extensively to unravel the pairing symmetry of the ground state of bulk superconductors. In conventional Bardeen-Cooper-Schrieffer (BCS) superconductors,
with a fully gapped excitation spectrum, the penetration depth exhibits exponential  behavior at low temperatures. In contrast, in unconventional superconductors, with nodal lines or nodal points of the gap function on the Fermi surface, $\lambda(T)$ shows power-law behavior depending on the type of nodes.\cite{EinzelYEAR86,GrossYEAR86,BarashYEAR96,AnnetBook}
Therefore the observation of power laws in $\Delta\lambda(T) \equiv \lambda(T)-\lambda(0)$ has been taken synonymous with unconventional pairing symmetries in the heavy-fermion and high-temperature copper-oxide superconductors.

It is known that disorder changes the power-law behavior from linear to quadratic in $T$ for $d$-wave superconductivity as disorder fills in  impurity states in the nodal gap
regions.\cite{ChoiYEAR88,KlemmYEAR88,AnnetPRB,ProhammerYEAR91,ArbergYEAR93,HirschfeldYEAR93,XuYEAR95}
Since then it was shown that  disorder in unconventional superconductors leads to an even stronger suppression of the superfluid density $\rho_s$ relative to its superconducting transition temperature $T_c$ than expected from unitarity scattering in the Abrikosov-Gor'kov (AG) theory. In particular, Franz and co-workers\cite{FranzYEAR97} investigated short-coherence-length superconductivity in cuprates, that is, the coherence length $\xi$ is comparable to the lattice parameter $a$, and demonstrated that it behaves markedly differently from the AG theory of impurity-averaged order parameters.
In fact, detailed calculations of the spatial dependence of the local density of states and order parameter in the vicinity of an impurity with strong (unitarity) nonmagnetic scattering potential showed that the order parameter is abruptly suppressed
within just a few lattice parameters resembling the holes within a Swiss
cheese.\cite{ByersYEAR93,BalatskyYEAR95,FlatteYEAR97,SalkolaYEAR97,CSTing,BalatskyRMP}
This is in stark contrast to the assumption in the AG theory of pair-breaking with a  spatially uniform suppressed impurity-averaged order parameter. The Swiss Cheese model was originally introduced to explain the universal scaling behavior of superconducting transition temperature versus zero-temperature superfluid density in the Uemura plot, $T_c$ vs.\ $\rho_s(0)$,\cite{UemuraYEAR91} of underdoped high-$T_c$ cuprate superconductors\cite{NachumiYEAR96,UemuraYEAR00} and found a recent revival for describing the behavior of Kondo holes in heavy-fermion superconductors.\cite{BauerYEAR11}

In this work, we use the Bogoliubov-de Gennes (BdG)  lattice model to  study the effects of disorder on bulk and local properties in superconducting PuCoGa$_5$. Plutonium-based superconductors can be thought of as superconducting clocks, since the natural radioactivity of Pu ($^{239}$Pu half-life = $24,000$ years) creates lattice defects like clockwork, which scatter electrons and break Cooper pairs. Pair-breaking  creates impurity bands in the nodal regions affecting the signature of a pure $d$-wave superconductor. Such pair-breaking effects are evident in the suppression of $T_c$ with time,\cite{JutierYEAR2005,JutierYEAR05,JutierYEAR08} the spin-lattice relaxation rate
$1/T_1$,\cite{CurroYEAR05} where the $T^3$ temperature dependence switches to a linear-in-$T$ behavior, and the penetration depth $\lambda(T)$, where the linear-in-$T$ temperature dependence gives way to a $T^2$ behavior at low temperatures.\cite{OhishiYEAR07,MorrisYEAR06} Unlike other bulk probes, the magnitude of $\lambda(0)$ is very sensitive to  defects owing to its nature of measuring the stiffness of the superconducting phase coherence in the sample, whereas the magnitude of $T_c$ is less sensitive to defects, since it is related to the  spatial average of the order parameter. A detailed account of this difference will be presented.

The main result of our study is that self-irradiation-induced defects in  PuCoGa$_5$ violate  the AG theory of dilute disorder in superconductors, which is based on the premise of an impurity-averaged order parameter, while it is fully consistent with the Swiss Cheese model of short-coherence-length superconductors. Our detailed calculations show that measurements of the transition temperature and superfluid density  in fresh (25 day-old) and aged (400 day-old) PuCoGa$_5$ are fully consistent with the Swiss Cheese model of disorder, which captures both the 20\% suppression of the superconducting transition temperature $T_c$  and a large 70\% suppression of the superfluid density $\rho_{\rm s}(0)$  between aged and fresh samples, as well as the temperature dependence of \rhos. Finally, our study of the Swiss Cheese model exemplifies the common behavior of the superfluid density in unconventional short-coherence-length superconductors with disorder. Because of spatially extended quasiparticle excitations along the nodal directions of the order parameter, in the presence of a strong on-site impurity potential, one finds the suppression of \rhos\ to extend over several coherence lengths, whereas the suppression of the order parameter is very localized and limited to a few lattice sites. In that respect a close relationship between the copper-oxide and plutonium based superconductors exists.

The paper is organized as follows: In Sec.~II, we introduce the BdG lattice formalism of the Swiss Cheese model and the corresponding superfluid density expression. The local variation of the density of states,  superconducting order parameter, and superfluid density for a single impurity, as well as for multiple impurities, are presented in Sec.~III A. The evolution of the spatially averaged superfluid density and corresponding  density of states as a function of impurity concentration is discussed in Sec.~III B. In Sec.~III C we present the Uemura plot of $T_c$ vs.\ $\rho_s(0)$  for the Pu-based compound and compare with the results for cuprate superconductors.  Finally, we conclude in Sec.~V.

\section{Theoretical model}

We begin with the BdG mean-field theory of the attractive Hubbard model which was used extensively to describe superconductivity in correlated electron systems. Our main interest is focused on the linear response calculation of the superfluid density $\rho_{\rm s}$ for the BdG lattice model following the approach described by Scalapino {\it et al.}~\cite{ScalapinoYEAR92} for the Hubbard lattice model. This generic approach was consequently applied
to two-dimensional $d$-wave \cite{XiangYEAR95,TrivediYEAR96,FranzYEAR97}
and $s$-wave~\cite{XiangYEAR95,TrivediYEAR96,GhosalYEAR98,GhosalYEAR01,HurtYEAR05} superconductors.

\subsection{The BdG lattice model}

It has been shown that superconductivity in PuCoGa$_5$ is
unconventional because thermodynamic, transport, nuclear magnetic resonance, and neutron scattering data  are consistent
with a model based on $d$-wave pairing symmetry (that is, based on nodal lines in the gap function).\cite{SarraoYEAR02,CurroYEAR05,JutierYEAR08,HiessYEAR07,HiessYEAR08}
First-principles calculations\cite{HottaYEAR03,OpahleYEAR03,OpahleYEAR04,TanakaYEAR04,ShickYEAR05,HottaYEAR06,OppeneerYEAR07,ShickYEAR11}
have shown that PuCoGa$_5$ hosts a Fermi surface with four sheets, of which two are cylindrical sheets centered at the M point in the Brillouin zone. The quasi-two-dimensional nature is related to the layered structure of Pu atoms forming a square lattice.
For simplicity, we consider only a single-band model, which captures the essential physics of $d$-wave superconductors.
In order to describe the disorder effect on  superconductivity, we consider a tight-binding model Hamiltonian defined on a square lattice:
\begin{eqnarray}
H_0 &=& -\sum_{ij\sigma}  t_{ij} c_{i\sigma}^{\dagger} c_{j\sigma}
+ \sum_{i\sigma} (\epsilon_i - \mu) c_{i\sigma}^{\dagger} c_{i\sigma}
\nonumber\\
&&
+\sum_{ij}  \Delta_{ij} c_{i\uparrow}^{\dagger} c_{j\downarrow}^{\dagger}
+ \text{H.c.}
\label{EQ:Hamil}
\end{eqnarray}
Here $c_{i\sigma}^{\dagger}$ ($c_{i\sigma}$) creates (annihilates) an electron at the $i$-th site
of spin $\sigma$. The variables  $t_{ij}$ and $\mu$ are the hopping integrals and chemical
potential, respectively. We model the disorder by considering a distribution of short-ranged
nonmagnetic impurities, that is, $\epsilon_i=U_{\text{imp}}\delta_{iI}$, with $U_{\text{imp}}$
representing the potential scattering strength. The quantity $\Delta_{ij}$ denotes the superconducting order parameter
or gap function. Since the origin of superconductivity is not our concern,
we introduce an effective nearest-neighbor pairing interaction $V$, such that the superconducting order parameter is determined self-consistently:
\begin{equation}
\Delta_{ij} = \frac{V}{2} \langle c_{i\uparrow}c_{j\downarrow} - c_{i\downarrow}c_{j\uparrow} \rangle\;,
\end{equation}
where $(ij)$ is a nearest-neighbor (NN) site pair, and zero otherwise.
By using the Bogoliubov transformation
\begin{subequations}
\begin{eqnarray}
c_{i\uparrow} &=& \sum_{n} [u_{i}^{n} \gamma_{n} - v_{i}^{n*} \gamma_{n}^{\dagger}] \;, \\
c_{i\downarrow} &=& \sum_{n} [u_{i}^{n} \gamma_{n} + v_{i}^{n*} \gamma_{n}^{\dagger}] \;,
\end{eqnarray}
\end{subequations}
the  Hamiltonian in Eq.~(\ref{EQ:Hamil}) can be diagonalized
by solving the corresponding BdG equations\cite{CSTing,BdGbook}:
\begin{equation}
\sum_{j} \left(
\begin{array}{cc}
\mathcal{H}_{ij} & \Delta_{ij} \\
\Delta^{*}_{ij}  & - \mathcal{H}_{ij}^{*}
\end{array}
\right)
\left(
\begin{array}{c} u_{j}^{n} \\ v_{j}^{n} \end{array}
\right)
=E_{n} \left(
\begin{array}{c} u_{i}^{n} \\ v_{i}^{n} \end{array}
\right) \;.
\label{EQ:BdG}
\end{equation}
Here $(u_{i}^{n},v_{i}^{n})^{T}$ are the eigenfunctions at site $i$ corresponding to the
quasiparticle excitation energy $E_{n}$, and
the normal-state single-particle lattice Hamiltonian is
\begin{equation}
\mathcal{H}_{ij} = -t_{ij} + (\epsilon_{i}- \mu) \delta_{ij} \;.
\end{equation}
Throughout this work, we limit the hopping integrals only to the NN sites on the
square lattice, that is, $t_{i,i+\delta}=t$ for  $\delta=(\pm 1, 0)$ and $(0, \pm 1)$, and zero otherwise.
The self-consistency equation for the superconducting order parameter on the square
lattice is thus given by
\begin{equation}
\Delta_{i,j=i+\delta}= \frac{V}{2} \sum_{n} [ u_{i}^{n} v_{j}^{n*} + u_{j}^{n}v_{i}^{n*}]
\tanh \biggl{(}\frac{E_{n}}{2k_{B}T}\biggr{)}\;,
\label{EQ:OP}
\end{equation}
where $\delta=(\pm 1, 0)$ and $(0, \pm 1)$, and zero otherwise,
and the temperature is denoted by $T$. For $d_{x^2-y^2}$-wave pairing symmetry, the order
parameter along the $y$ direction has opposite sign compared to  the
$x$ direction, which is indeed obtained in the solutions.
Once the BdG equations are solved, many interesting properties can be explored. For example,
for a given concentration of disorder, the superconducting transition temperature can
be determined by the condition that the averaged superconducting order parameter
vanishes. Another important observable that can be tested experimentally is the local density of states (LDOS) at zero temperature,
which is given by
\begin{equation}
\rho_{i}(E) = 2\sum_{n} [\vert u_{i}^{n}\vert^2\delta(E-E_n) + \vert v_{i}^{n} \vert^2 \delta(E+E_n)] \;,
\end{equation}
where the prefactor `2' is due to the twofold spin degeneracy. The differential tunneling conductance
in scanning tunneling spectroscopy is directly proportional
to the LDOS and can provide insights into the electronic properties and symmetry of a superconductor.
Since one of the major pieces of interest in the present work is the superfluid
density, which characterizes the superconducting phase rigidity, it is useful to give
an expanded discussion of its derivation and lattice formulation. Our BdG lattice formulation of the superfluid density
follows closely the seminal work by Scalapino and coworkers\cite{ScalapinoYEAR92} for the Hubbard model on a lattice.
We calculate the superfluid stiffness for a current response to a vector potential of wave vector $\bm{q}$
and frequency $\omega$ along the $x$ direction
as given by the Kubo formula. For this purpose, we expand the Hamiltonian to include the interactions of electrons coupled to an electromagnetic field. The  tile-dependent total Hamiltonian is
\begin{equation}
H_{\text{t}} = H_0 + H^{\prime}(t)\;.
\end{equation}
Here $H^{\prime}(t)$ describes such a minimal coupling
\begin{equation}
H^{\prime}(t) = -e a \sum_{i} A_{x}(\mathbf{r}_i,t)
\left(
 J_{x}^{P}(\mathbf{r}_i) + \frac{e a}{2} A_{x}(\mathbf{r}_i,t)K_{x}(\mathbf{r}_i)
\right),
\end{equation}
where $a$ is the lattice constant,
$A_{x}$ is the vector potential along the $x$ axis, and
\begin{equation}
J_{x}^{P}(\mathbf{r}_i) = -i\sum_{\sigma,\delta} [t_{i,i+\delta} c_{i\sigma}^{\dagger}c_{i+\delta,\sigma} - \text{h.c.}]\;,
\end{equation}
\begin{equation}
K_{x}(\mathbf{r}_i) = -\sum_{\sigma,\delta} [t_{i,i+\delta} c_{i\sigma}^{\dagger}c_{i+\delta,\sigma} + \text{h.c.}]\;,
\end{equation}
are the particle current and kinetic energy densities. The variable $\delta =
\hat{x}, \hat{x}\pm \hat{y}$ denotes the links which have contribution to
the bond current and kinetic energy along the $x$ axis.
The charge current density operator along the $x$ axis is then found to be
\begin{equation}
J_{x}^{Q}(\mathbf{r}_i) \equiv - \frac{\delta H^{\prime}(t)}{\delta A_{x}(\mathbf{r}_i,t)}
= eJ_{x}^{P}(\mathbf{r}_i) + e^{2} K_{x}(\mathbf{r}_i) A_{x}(\mathbf{r}_{i},t)\;.
\end{equation}
We calculate the paramagnetic component of the electric current density
to first order in $A_x$,
\begin{equation}
\langle J_{x}^{P}(\mathbf{r}_i)\rangle = -i \int_{-\infty}^{t} \langle [J_{x}^{P}(t),H^{\prime}(t^{\prime})]_{-}\rangle_{0} dt^{\prime} ,
\end{equation}
and the diamagnetic part in $\langle K_{x} \rangle_{0}$ only to zeroth order;
$\langle \dots \rangle_0$ represents a thermodynamic average with respect to $H_0$.
Straightforward algebra yields the current response function
\begin{eqnarray}
-\frac{J_{x}^{Q}(\mathbf{r}_i)}{e^{2}A_{x}(\mathbf{r}_{i})} &=& -
ie^{-i\mathbf{q}\cdot \mathbf{r}_i} \int_{-\infty}^{t} dt^{\prime}
 \langle[J_{x}^{P}(\mathbf{q},t),J_{x}^{P}(-\mathbf{q},t^{\prime})]_{-}\rangle_0\nonumber\\
&&  -\langle K_{x}(\mathbf{r}_i) \rangle_0\;.
\end{eqnarray}
By performing a lattice average over the variable $\mathbf{r}_i$ to eliminate
the atomic-scale fluctuations, we define an effective ``Drude weight'' as a measure
of
the superfluid density
\begin{equation}
\rho_s \equiv \frac{D_s}{\pi e^{2}} = -\langle K_{x} \rangle + \Pi_{xx}(\mathbf{q}\rightarrow 0,\omega=0)\;.
\end{equation}
The first term is the kinetic energy along the $x$ direction divided by the number of lattice sites,
\begin{eqnarray}
\langle K_{x} \rangle &=& -\frac{t}{N} \sum_{i,n,\sigma} \Big[ f(E_{n}) [u_{i+x,\sigma}^{n*} u_{i\sigma}^{n} + \text{c.c.} ]\nonumber \\
&& + (1-f(E_{n}))[v_{i+x,\sigma}^{n} v_{i\sigma}^{n*} + \text{c.c}]\Big]\;.
\end{eqnarray}
It represents the diamagnetic response to an external magnetic field $\bm{B}=\nabla \times \bm{A}$ with gauge $A_x \neq 0$ and $A_y=A_z =0$.
The second term is the paramagnetic response given by the disorder-averaged transverse current-current correlation function
\begin{widetext}
\begin{equation}
\Pi_{xx}(\mathbf{q}\rightarrow 0 ) = \frac{1}{N} \sum_{n_1,n_2} \biggl{\{}
\frac{A_{n_1,n_2}(\mathbf{q}\rightarrow 0 )[A_{n_1,n_2}^{*}(\mathbf{q}\rightarrow 0) + D_{n_1,n_2}(-\mathbf{q}\rightarrow 0)]}{E_{n_1}-E_{n_2} }  [f(E_{n_1})-f(E_{n_2})]\biggr{\}} \;,
\end{equation}
\end{widetext}
with coefficients
\begin{eqnarray}
A_{n_1,n_2}(\mathbf{q}) &=& 2\sum_{i} e^{-i\mathbf{q}\cdot \mathbf{r}_{i}}[u_{i+\hat{x}}^{n_1 *} u_{i}^{n_2} -
u_{i}^{n_1 *} u_{i+\hat{x}}^{n_2} ]\;, \\
D_{n_1,n_2}(\mathbf{q}) &=& 2 \sum_{i} e^{-i\mathbf{q} \cdot \mathbf{r}_{i}} [v_{i+\hat{x}}^{n_1}
v_{i}^{n_2 *} - v_{i}^{n_1} v_{i+\hat{x}}^{n_2 *}]\;.
\end{eqnarray}

\subsection{Numerical solution of the BdG equations}
\label{BdG_parameters}

We follow an iterative numerical procedure to solve self-consistently
the BdG equations  via exact diagonalization: For a given impurity distribution, we start
with a uniform $d_{x^2-y^2}$-wave order parameter, that is $\Delta_{i,i+\hat{x}} = -\Delta_{i,i+\hat{y}}$,
at a low temperature. After the BdG equations~(\ref{EQ:BdG})
are diagonalized, the obtained eigenvalues and  eigenfunctions are used to update the superconducting order parameter as given by Eq.~(\ref{EQ:OP}). Then we start a new cycle of iteration. The iteration will continue until a convergence criterion for the order parameter is reached. We have taken the difference of the order parameter at all sites between two
consecutive iterations to be less than $10^{-8}$ as the convergence criterion.
Upon convergence is reached at a given temperature, both the local density of states (LDOS) and superfluid density are calculated by using the supercell technique to reduce finite-size effects. The final LDOS and superfluid density are the result of an ensemble average over about 20 impurity configurations. After this step, the calculation is moved to the next higher temperature point.

Next we also comment on the choice of lattice model parameters.
In our numerical calculations, we have chosen to measure the energy
in units of the nearest-neighbor hopping parameter $t$ with chemical potential $\mu=-0.36t$ and superconducting pairing interaction strength $V=1.13t$. The on-site impurity potential scattering strength $U_{\text{imp}}=100t$
was used to model the strong (unitarity) limit of impurity scattering. Additionally, we broadened the density of states calculations by a small imaginary term of width $\Gamma=0.01 t$ to overcome the discreteness of the energy spectrum.
The size of a single cell for the self-consistent lattice model calculation is at least of 20-by-20 sites, when the supercell method with 6-by-6 cells is employed or 35-by-35 sites otherwise. However, we still encountered finite size effects for systems of such size at low temperatures and for low disorder concentrations.

\section{Results and discussion}
\subsection{Real-space imaging of local properties around a single impurity}
\begin{figure*}
\rotatebox[origin=c]{0}{\includegraphics[width=2.\columnwidth]{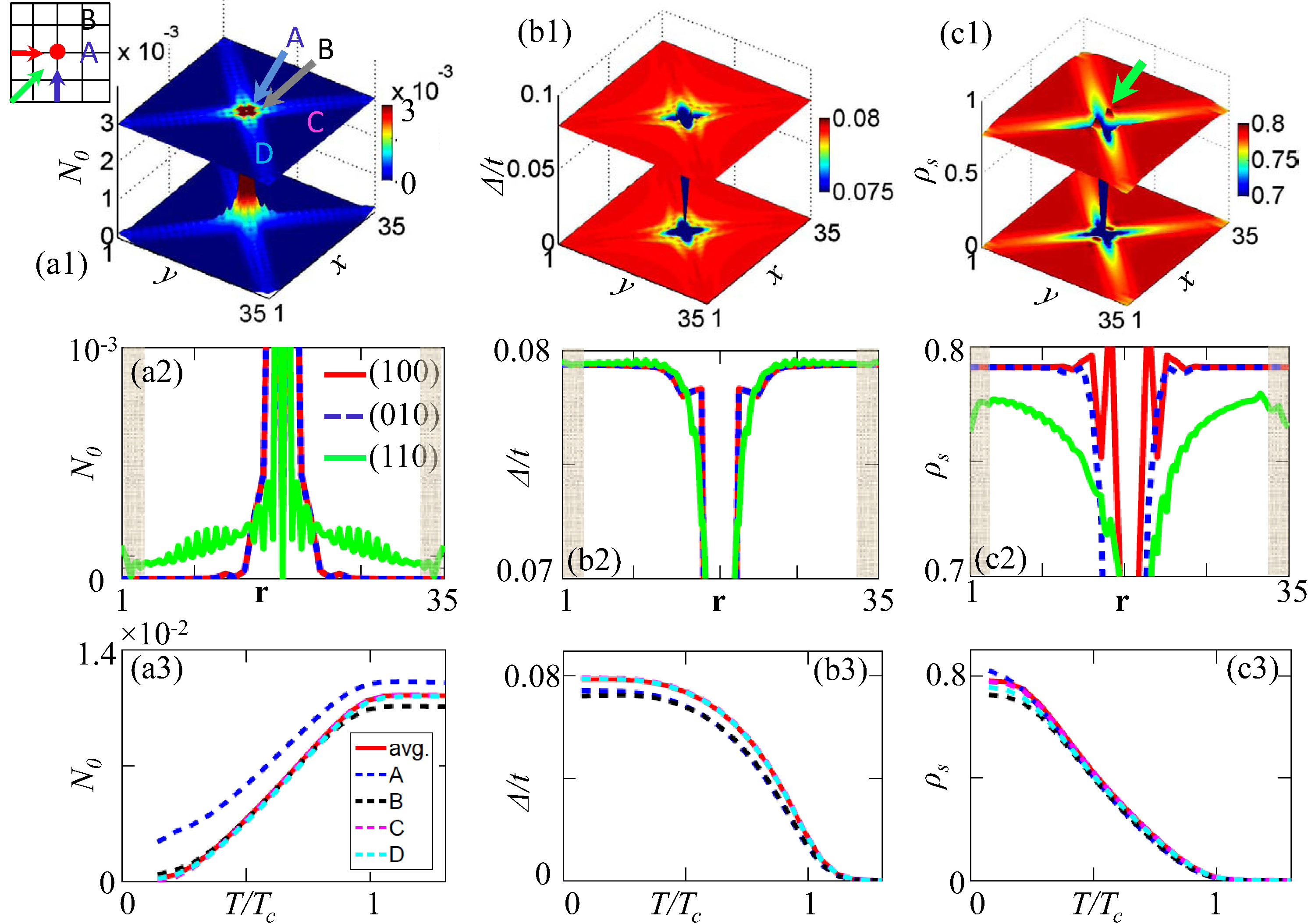}}
 \caption{(Color online) Visualization of the local behavior of superconducting properties for a single impurity at the center of the cell, see panels (a1)-(c1).
The 3D plots of $N_0({\bf r})$ in (a1), $\Delta({\bf r})$ in (b1) and $\rho_{\rm s}({\bf r})$ in (c1) show the real-space  modulation of these properties in response to a single impurity at the center. The corresponding color-coded contour maps highlight the patterns of the modulations. All spectra are calculated on a 35-by-35 lattice and then interpolated for visualization. The {\it inset} to (a1) gives the schematic view of various directions and locations of sites with respect to an impurity (red dot) at which the following plots are drawn. Each plot in the middle row corresponds to a 1D cut through the spectrum shown in the corresponding top panel (the color of each representative curve is the same as for the arrow drawn in the {\it inset}). The shading at the lattice boundaries delineates the region where finite-size effects are expected to affect the results shown in panels (a3)-(c3). Similarly the curves in the bottom row are drawn at four representative sites with respect to the location of the impurity and are compared with their average value. Site `A' is the nearest neighbor to the impurity along the antinodal direction. The next-nearest-neighbor site `B' is along the nodal direction. Sites `C' and `D' are farther away from the impurity location along the antinodal and nodal directions, respectively.} \label{rho_1imp}
\end{figure*}

We begin with spatial images of various local properties at $T=0$, including the LDOS at zero excitation energy $N_0({\bf r})$,  the superconducting (SC) gap amplitude $|\Delta({\bf r})|$, and the superfluid density $\rho_{{\rm s}}({\bf r})$ all shown in the top row of Fig.~1 for a single impurity at the center of the cell. The SC gap amplitude at a particular site $(x,y)$ is obtained from $|\Delta({\bf r})|=\big[\Delta(x+\delta x,y)+\Delta(x-\delta x,y)-\Delta(x,y+\delta y)-\Delta(x,y-\delta y)\big]/4$, where $\delta$ is the distance between NN sites. [Below we denote $\Delta(0)$ as the average SC gap at $T=0$.] Whereas $\rho_{\rm s}({\bf r})$ is the current-current correlation function which involves a double summation over real-space. Inserting Eqs.~(18)-(19) into Eq.~18, we obtain $\rho_{\rm s}({\bf r})\sim\sum_{i,j}A^i(A^j+D^j)$ with site indices ($i, j$). We plot $\rho_{\rm s}({\bf r})$ in Fig.~1(c1) to visualize the effect of a single impurity on the rigidity of the superconducting phase.

The well-known Friedel oscillation is clearly seen in the LDOS $N_0({\bf r})$ along the nodal direction of the SC gap. This is a signature of unconventional superconductors, which was studied extensively in the high-$T_c$
cuprates\cite{ByersYEAR93,BalatskyYEAR95,FlatteYEAR97,SalkolaYEAR97,FranzYEAR97,BalatskyRMP} and iron-based superconductors.\cite{Hirschfeld,JianXin} The $d_{x^2-y^2}$-wave order parameter is fully suppressed at the impurity site for a strong scatterer, aside from very weak oscillations near the order parameter maximum, see Fig.~1(b1). The spatial dependence of the superfluid density $\rho_{{\rm s}}({\bf r})$ is also strongly localized around the impurity with characteristic features more similar to the LDOS than the order parameter. To illustrate this type of Swiss cheese phenomenon of strongly suppressed superconducting properties near an impurity, we have taken various cuts along (100), (010), and (110) directions, as shown by solid red, dashed blue and solid green lines, respectively, in the middle row in Fig.~1.

Next we revisit characteristic properties of a single impurity in a $d$-wave superconductor.
The LDOS is a measure of the single-quasiparticle spectral weight. In agreement with earlier works\cite{ByersYEAR93,BalatskyYEAR95,FlatteYEAR97,SalkolaYEAR97,BalatskyRMP}  it decays inversely quadratic with distance $r$ from the impurity, $1/r^{2}$, along the nodal direction of the gap (green line), whereas along the antinodal direction, the decay is exponential. The spatial order parameter amplitude $|\Delta({\bf r})|$, plotted in Fig.~1(b2), creates a resonance state at the impurity site, reflecting the pair-breaking characteristics of $d$-wave pairing. The fourfold modulation of the order parameter is preserved in the case of a scalar impurity, while the suppression is smoother along the nodal direction where strongly damped Friedel oscillations are found. In contrast, the superfluid density $\rho_{{\rm s}}({\bf r})$ shows the usual fourfold modulation but in addition picks up the phase of the $d_{x^2-y^2}$-wave pairing symmetry as shown in Fig.~1(c2). Along  the (100)-antinodal direction, where the gap is positive, $\rho_{{\rm s}}$ shows a remarkable enhancement from its average value at the NN lattice site from the impurity (red line), whereas the NN lattice site along the (010)-antinodal direction exhibits stronger suppression (blue dashed line). This result is expected in linear response for a supercurrent flowing along the (100) direction, because it breaks the tetragonal symmetry of the lattice. Along the diagonal direction, the power-law behavior of $\rho_{{\rm s}}$ arises from gapless quasiparticles in the LDOS $N_0({\bf r})$.\cite{ByersYEAR93,BalatskyYEAR95,FlatteYEAR97,SalkolaYEAR97,CSTing,BalatskyRMP}

To gain further insight into the impurity effect on $\rho_{\rm s}({\bf r})$, we plot the temperature evolution of these quantities at four representative sites with respect to the impurity position and compare them with the average value. In the case of a clean superconductor with lines of nodes in the gap function, the low-temperature ($k_BT\ll \Delta$) approximation yields a linear behavior of the superfluid  density for gapless nodal quasiparticles on the Fermi
surface: \cite{ChoiYEAR88,KlemmYEAR88,AnnetPRB,ProhammerYEAR91,ArbergYEAR93,HirschfeldYEAR93,XuYEAR95,DurstYEAR00,Das_rho}
%
\begin{equation}
\rho_{\rm s}(T) / \rho_{\rm s}(0) \simeq 1 - C_1T.
\end{equation}
Here, the slope $C_1$ is determined by the BCS ratio $\Delta(0)/T_c$. In the presence of an impurity with large broadening $\gg k_BT$, the lowest order temperature dependence of the superfluid density becomes $T^2$ [Refs.~\onlinecite{AnnetPRB,ProhammerYEAR91,ArbergYEAR93,HirschfeldYEAR93,XuYEAR95}] (note that for an isotropic $s$-wave gap the low-energy superfluid density decays exponentially):
\begin{equation}
\rho_{\rm s}(T)/ \rho_{\rm s}(0) \simeq  1 -C_2T^2.
\end{equation}
Here ${C_2}$ is a more complex function of Fermi velocity and effective impurity broadening.\cite{HirschfeldYEAR93} In Fig.~1(c3) we see that at the NN site `A' along the (100) direction, where $\rho_{{\rm s}}(A)$ shows a peak, the corresponding  $\rho_{{\rm s}}(A,T)$ shows a quasilinear behavior at low temperatures, within the error produced by finite-size effect. At the NNN site `B' along the node, the superfluid density is suppressed showing a $T^2$ dependence in accord with the average value. We believe that in principle magnetic field-angle dependent measurements of the magnetization can provide a unique and indispensable tool to probe the presence of gap nodes and shed light on the pairing symmetry in the ground state of a bulk superconductor.
This idea is related to the observation of the nonlinear Meissner effect predicted by Yip and Sauls\cite{YipYEAR92,XuYEAR95} for fields along the nodal and antinodal directions. However, its analysis will be complicated by surface bound states.\cite{WalterYEAR99}

\begin{figure*}
  \rotatebox[origin=c]{0}{\includegraphics[width=2.\columnwidth]{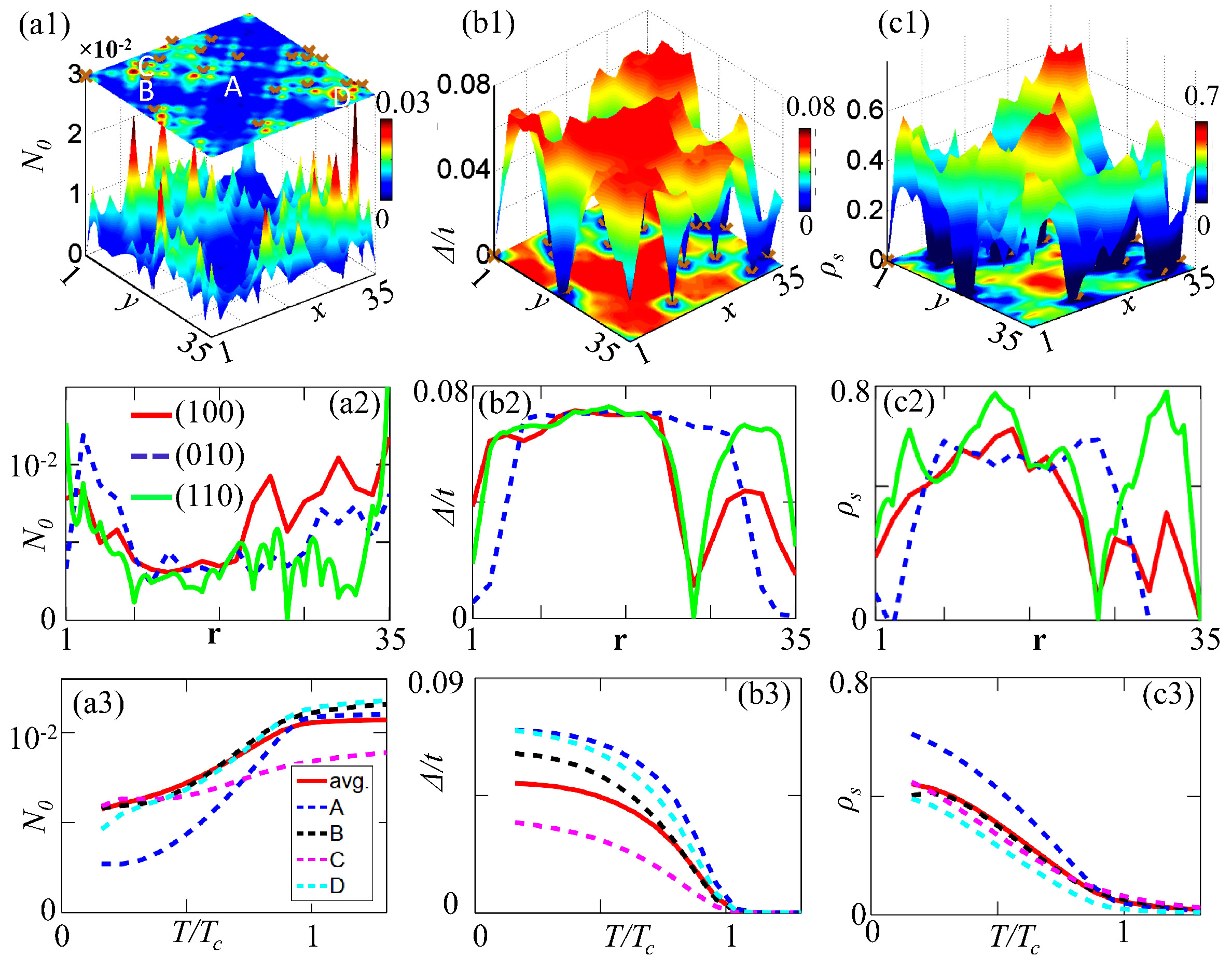}}
 \caption{(Color online) Visualization of the local behavior of superconducting properties for a randomly distributed impurity concentration of 2\%, same notation as in Fig.~1. Brown cross symbols on the 2D color maps in the top row depict the locations of impurities on the lattice. The selection of test sites used in the bottom row of panels is the same for all spectra and marked in panel (a1). Unlike for the single-impurity case in Fig.~1, the SC gap and superfluid density now show much stronger spatial fluctuations. All averaged calculations are done for 10 samplings over random configurations of impurities. The  visualized results are for one particular disorder configuration.} \label{rho_2imp}
\end{figure*}

Next we turn to the case of randomly distributed impurities in Fig.~2 for a concentration $n_{imp}=2\%$, where the local information of the previously discussed quantities changes more dramatically across the entire lattice. Nevertheless, the local response of all quantities with respect to the impurity location as well as the one-to-one correspondence between LDOS, order parameter, and superfluid density is present for all impurity concentrations studied. Differences between multiple impurities and single impurity are appreciable in all calculated spectra. Due to the quasiparticle interference and the overlap of wavefunctions of quasiparticles scattering off impurities, Friedel oscillations are enhanced and present for all directions shown in Fig.~2(a2)-(c2).
Similarly, the differences in the $T$ dependence of properties at each site is more clearly visible here as can be seen in Fig.~2(a3)-(c3). For example, at position `B', which sits at the center along the nodal direction between two impurities, one can probe the gapless quasiparticles and thus show quasi-linear behavior in $\rho_{\rm s}(T)$ at low $T$. Whereas at site `A', which sits nearly at the center between two impurities, but along an off-nodal direction, one sees enhancement in $\Delta(0)$ and $\rho_{\rm s}$ at low $T$.

\subsection{Temperature dependence of average superfluid density}

\begin{figure*}
\rotatebox[origin=c]{0}{\includegraphics[width=1.95\columnwidth]{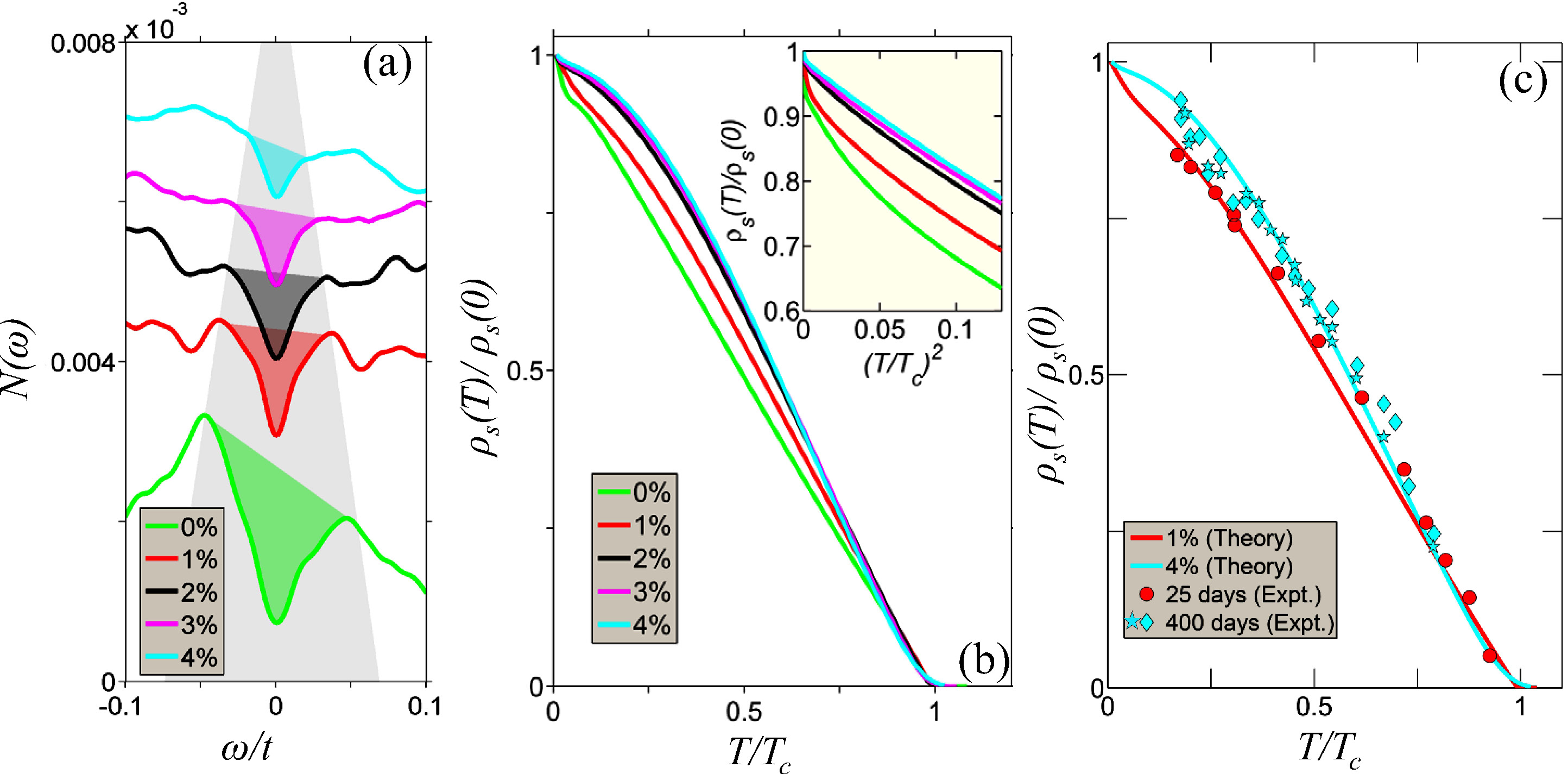}}
 \caption{(Color online) The energy dependence of the DOS is related to the temperature dependence of $\rho_{\rm s}(T)$ (averaged over 35-by-35 lattice) for different impurity concentrations.
(a) Computed DOS $N(\omega)$ is plotted for increasing impurity concentration along the vertical axis (curves are not shifted vertically). The grey shaded area highlights the nature of the gap closing, while the colored filling in each spectrum illustrates the trend of gap filling with increasing disorder. Consequences of finite-size effects can be seen in the value of the zero-energy DOS for the pure sample (red line), which is expected to be zero in the absence of a numerical broadening term $\Gamma$.
(b) Calculated results of $\rho_{\rm s}(T)$ (normalized to their corresponding zero temperature value). The {\it inset} expands the low-$T$ region of $\rho_{\rm s}(T)$ vs.\ $T^2$ to emphasize the quadratic temperature dependence.
(c) Computed $\rho_{\rm s}(T)$ for 1\% and 4\% impurity concentrations are compared with the measured data for a fresh (25 day-old) and old (400 day-old) PuCoGa$_5$ sample, respectively.\cite{OhishiYEAR07} The cyan diamond and red circle symbols are for field-cooled measurements in $H_0 = 60$ mT, while the cyan star symbols are for $H_0 = 300$ mT.
Since no data are available at zero temperature and near $T_c$, we extrapolated each experimental curve into these regions for proper normalization. After extrapolating the data, we estimated the values of $\rho_s(0)$ and $T_c$, which are slightly higher than the ones predicted using a linear extrapolation method;\cite{OhishiYEAR07,JutierYEAR08,CurroYEAR05} see Table I. } \label{rho_T}
\end{figure*}

Figure~3 presents the calculated DOS (averaged over the entire 35-by-35 lattice sites) as a function of energy for the $d$-wave pairing case and impurity concentrations $n_{imp}=0, 1, 2, 3, 4$\%.
The disorder-induced change in the DOS is also reflected in the temperature dependence of the spatially averaged superfluid density. Since finite-size effects are more pronounced at low energy and for low impurity concentrations, for example, the zero-energy DOS of the pure sample (red line) in Fig.~3(a) is expected to vanish, while it is of the order of the numerical broadening term $\sim\sqrt{\Gamma/\Delta(0)} \,N_n(0)$, where $N_n(0)$ is the normal state DOS at the Fermi level. For the same reasons, the low-temperature values of $\rho_{\rm s}(T)$ are expected to be less accurate. Nevertheless, our calculations reproduce the hallmark `V'-shaped feature of the DOS due to $d-$wave pairing for all concentrations. It is interesting to point out that, with increasing concentration not only is the gap amplitude decreasing, but also the gap nodes become more filled in by impurity states, see shaded and colored areas in Fig.~3(a). The gap-filling or build-up of resonant impurity states near the Fermi level (which can be quantified by an impurity scattering rate) are the main aspect of impurity effects that give rise to the quadratic-in-$T$ dependence of the superfluid density at low temperatures. For large impurity concentrations the Swiss Cheese model recovers the dirty $d$-wave result of  Eq.~(21) for the temperature dependence of \rhos. Plotting $\rho_{\rm s}(T)$ as a function of $T^2$ in the {\it inset} of Fig.~3(b) demonstrates the gradual change from linear to quadratic behavior with increasing impurity concentration.

To connect our Swiss Cheese model calculations with muon-spin rotation measurements of the penetration depth,
we compare in detail in Fig.~3(c) the temperature dependence of \rhos\ of the fresh and aged samples with our BdG calculations in the limit of  strong on-site impurity scattering. The impurity concentrations $n_{\text{imp}} = 1\%$ and  $4\%$ were chosen to reproduce the observed suppression of $T_c$ and $\rho_{\rm s}(0)$ of the fresh and aged samples, respectively.  The relevant model parameters were given in Sec.~\ref{BdG_parameters}, which are related to the coherence length $\xi \sim \hbar v_F/(\pi\Delta)$, where the Fermi velocity is given approximately by
$v_F \sim a t/\hbar$ with $\Delta=4\Delta(0)$ and $\Delta(0) = 0.08t$ at $n_{imp}=4\%$,
so that $\xi \sim 3.9a$ and lattice parameter $a=0.423$ nm.\cite{SarraoYEAR02} From our calculated $T_c$ suppression and comparison with experiment, we conclude that hypothetically pure PuCoGa$_5$ has a bare superconducting transition of $T_{c0}=18.9$ K in agreement with previous estimates.
The reported decrease in $T_c$ of about 3 K is in agreement with the radiation-induced reduction of $T_c$ ($\approx
0.24$ K/month) reported for PuCoGa$_5$ samples of slightly different isotopic concentrations.\cite{CurroYEAR05,JutierYEAR05,JutierYEAR08} We list in Table~I all characteristic parameters. Measurements on the fresh sample (after 25 days) were performed with an applied field $H_0 = 60$ mT. The aged sample (after 400 days) was measured in applied fields of 60 mT and 300 mT. All measurements were performed in a field-cooled mode above the lower critical field $H_{c1}$.\cite{OhishiYEAR07}

The excellent agreement between the short-coherence-length BdG calculations within the Swiss Cheese model and the
measured $T_c$ and superfluid density \rhos, combined with the previously demonstrated failure of the dirty $d$-wave theory,\cite{OhishiYEAR07}
shows that the uniform, dilute-impurity pair-breaking theory by Abrikosov and Gor'kov is not applicable to PuCoGa$_5$.

\begin{table*}
\caption{Parameters derived from comparing $\lambda(T)$ in
 25 day-old (fresh) and 400 day-old (aged) PuCoGa$_5$ to the BdG lattice model within
 the Swiss Cheese model of $d_{x^2-y^2}$-wave superconductivity, penetration depth
 $\lambda(0)$ (at $T = 0$), impurity concentration $n_{imp}$, and the
 transition temperature of the nominally pure sample $T_{c0}\simeq 18.9$ K. }
\begin{ruledtabular}
\begin{tabular}{cccccc}
Samples               & $n_{imp}$ & $T_c$ in K &    $T_c$ in K & $\lambda(0)$ (nm) & $\lambda(0)$ (nm) \\
             & [Present Theory]  & [Present calculation] &    [From Refs.~\onlinecite{JutierYEAR08,CurroYEAR05}] &  [Present calculation] & [From Refs.~\onlinecite{JutierYEAR08,CurroYEAR05}] \\
\\ \hline
Fresh (25 days)     & 1\%                       &   18.88      &    18.25(10)    &    310                    &  265(5)                    \\
Aged (400 days)   & 4\%                        &   16.45      &    15.0(1) 	  &   524                    &  498(10)                             \\
\end{tabular}
\end{ruledtabular}
\end{table*}

\subsection{Suppression of $\rho_{\rm s}(0)$ and $T_c$}

\begin{figure}[tp]
\rotatebox[origin=c]{0}{\includegraphics[width=0.95\columnwidth]{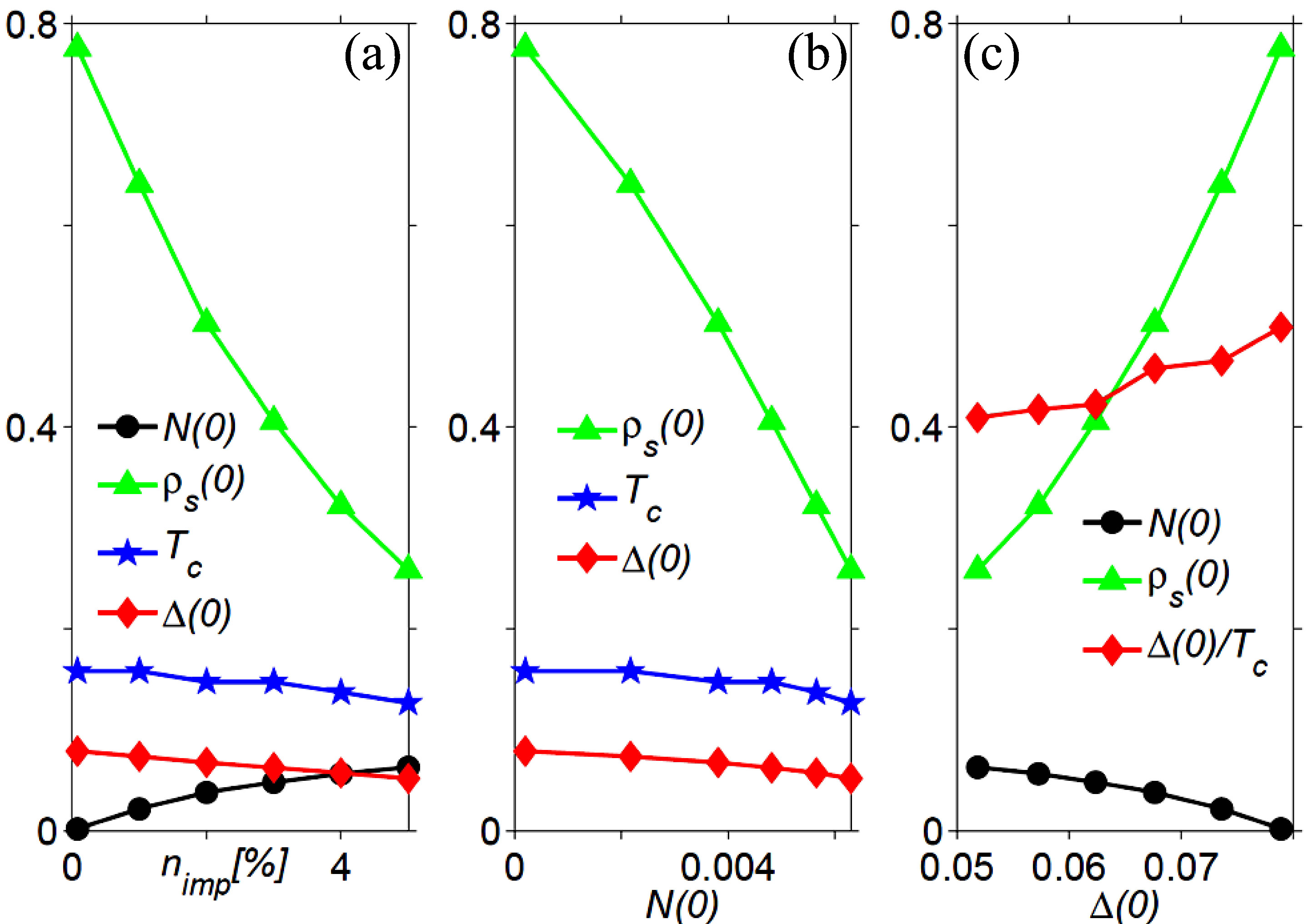}}
\caption{(Color online) Disorder induced correlations between $N(0)$, $\Delta(0)$, $T_c$ and $\rho_{\rm s}(0)$. (a) Variation of different properties (see legend) as a function of impurity concentration $n_{imp}$\%. (b)-(c) The same quantities are drawn as a function of total number of quasiparticles on the Fermi surface in (b) and as a function of order parameter suppression. }
\label{rho_nimp}
\end{figure}

The magnitudes of the superconducting properties $N(0)$, $T_c$ and $\rho_{\rm s}(0)$
provide valuable information about the topology of the gap function on the Fermi surface as a function of impurity concentration, see Fig.~4(a). Furthermore, the correlation between them exemplifies the deviations from conventional dirty $d$-wave theory arising in a system with short-coherence-length superconductivity.
While the dependence of $N(0)$ and $\Delta(0)$ (also $T_c$) on disorder is quasilinear, the suppression of the superfluid density is dramatically enhanced. These features demonstrate the strong deviation of the Swiss Cheese model from conventional AG theory as discussed in details below. It has been noted before that the suppression of the zero-temperature superfluid density for $d$-wave pairing with disorder on a square lattice is much stronger than for the superconducting gap or transition temperature.\cite{XiangYEAR95,FranzYEAR97} In Fig.~4(b) we show that $\Delta(0)$ and $T_c$ exhibit a weakly linear decrease with an increasing DOS at the Fermi level, $N(0)$. On the other hand, $\rho_{\rm s}(0)$ decreases faster than linear with increasing $N(0)$. The superfluid phase coherence is destroyed more rapidly than the superconducting amplitude accounting for the marked difference between fresh and aged samples. The dependence of the superfluid density on the order parameter is given in Fig.~4(c), while the same information as a function of $T_c$ is presented in the Uemura plot in Fig.~5. The BCS ratio $\Delta(0)/T_c$ increases gradually with increasing gap amplitude, that is, with decreasing impurity concentration as expected. As mentioned before, $\rho_{\rm s}$ is suppressed much faster than $T_c$ or $\Delta(0)$
For $n_{imp}=4$\% we calculate for $T_c$ or $\Delta(0)$ a suppression of $\sim 20$\%, whereas $\rho_{\rm s}(0)$ is suppressed by almost 70\%. This finding is consistent with the experimental data for cuprates where the superconducting transition temperature is much more robust to disorder than what is predicted by the AG theory for $d$-wave pairing, when measured against the corresponding change in $\rho_{\rm s}(0)$.

\begin{figure}[top]
\rotatebox[origin=c]{0}{\includegraphics[width=1.0\columnwidth]{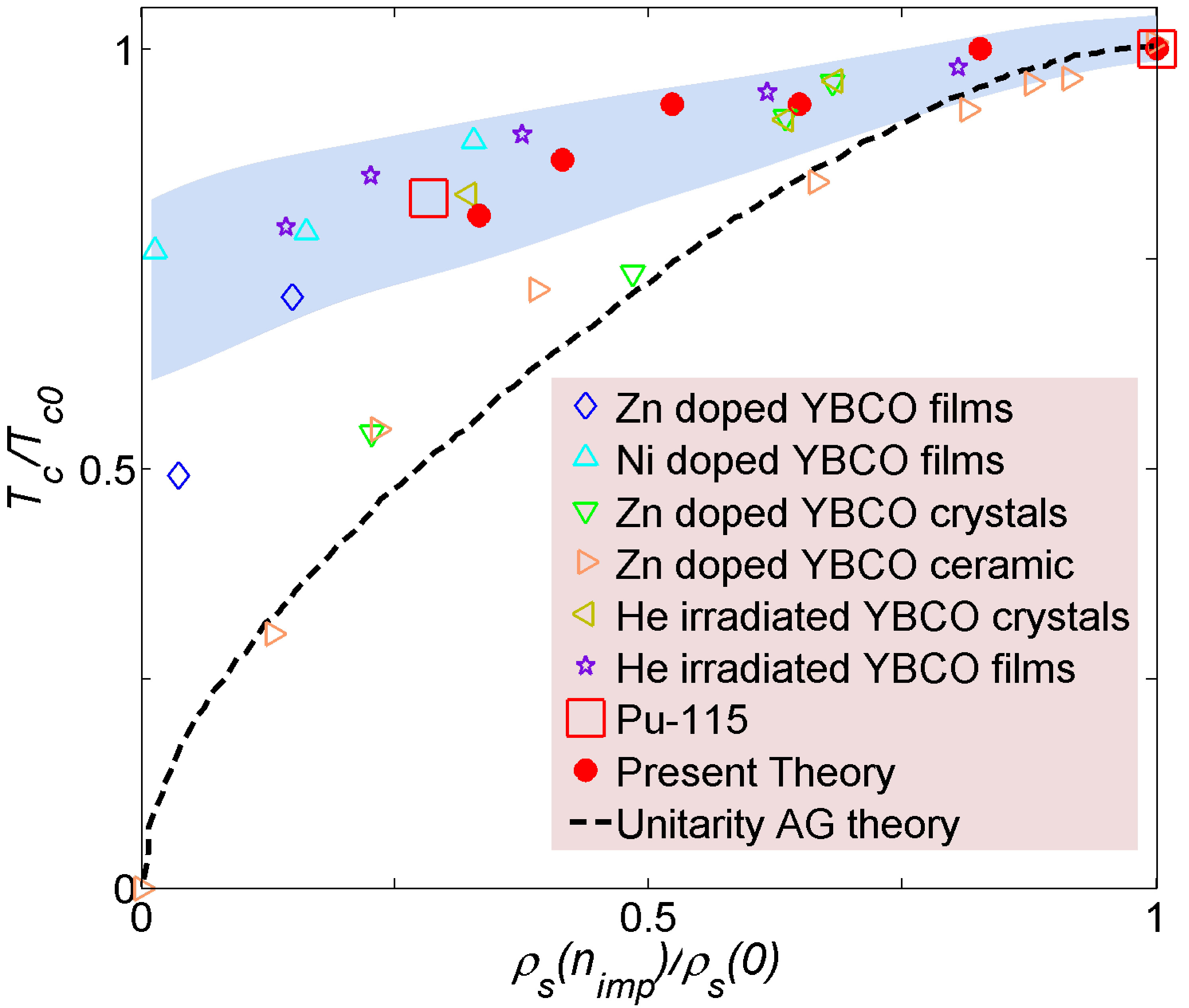}}
\caption{(Color online) Uemura plot of superfluid density in a disordered system. The two red open squares are the experimental data for PuCoGa$_5$ (Pu-115) from Ref.~\onlinecite{OhishiYEAR07}. The rest of the open symbols are taken from YBCO data under different conditions of disorder environments.\cite{FranzYEAR97} Data for YBCO films are obtained from Ref.~\onlinecite{Ulm}, for YBCO crystals from Ref.~\onlinecite{Basov}, for ceramic samples from Ref.~\onlinecite{Bernhard} and for He-irradiated YBCO films from Ref.~\onlinecite{Moffat}. The filled red dots are the present BdG theory. }
\label{rho_Tc}
\end{figure}

To elaborate some more on this point, we plot the experimental data of $\rho_{\rm s}(0)$ vs.\ $T_c$ in Fig.~5 for PuCoGa$_5$ by Ohishi {\it et al}.\cite{OhishiYEAR07} and compare with available experimental data on YBa$_2$Cu$_3$O$_{7-\delta}$ (YBCO) superconductors, the self-consistent T-matrix result of the dirty $d$-wave theory,\cite{PreostiYEAR94} and the Swiss Cheese model of the BdG lattice model. The situation of PuCoGa$_5$ is not unprecedented. For example, in YBCO $\rho_{\rm s}(0)$ is suppressed much more dramatically than $T_c$.
In Ni-doped and He-irradiated YBCO, a suppression of $T_c$ by about 20\% is accompanied by a suppression of $\rho_{\rm s}(0)$ by about 70\%.\cite{Ulm,Basov,MoffatYEAR97,NachumiYEAR96}
Several years ago, the same underlying disorder physics has been discussed within the BdG lattice model.\cite{FranzYEAR97,ZhitomirskyYEAR98,HettlerYEAR99}
Where applicable we find agreement with these calculations.
Franz {\it et al}.\cite{FranzYEAR97} compared their BdG results to YBCO samples for various conditions of disorder\cite{Ulm,Basov,Bernhard,Moffat} and concluded that the effect is enhanced by a short coherence length $\xi/a\approx$2-5.\cite{FranzYEAR97} As in YBCO, the coherence
length in PuCoGa$_5$ is relatively small ($\sim 2$ nm in both materials) with comparable lattice parameters ($\sim 0.5$ nm in both materials). Note that a similar enhanced suppression of the superfluid density with chemical doping has been reported for the related cerium-based compound CeCoIn$_{5-x}$Sn$_x$, where $\xi \approx 3$ nm.\cite{BauerYEAR06}

Taking into account the considerable spread among all data sets, a fit gives $T_c/T_{c0}\propto[\rho_{\rm s}(n_{imp})/\rho_{\rm s}(0)]^{0.4}$, which deviates significantly from the linear scaling of the Uemura plot ($T_c \propto \rho_{\rm s}(0)$) of underdoped high-$T_c$ cuprates, which has its origin in strongly correlated electron interactions.
Finally, it is evident that the AG theory overestimates the suppression of $T_c$, while the Swiss Cheese model is in very good agreement with all data sets.

\section{Conclusions}

We have found good agreement between the results of the Swiss Cheese model using the BdG lattice model and the superconducting properties of PuCoGa$_5$. Most importantly, the results demonstrate that, despite strong electronic inhomogeneity, a one-to-one correspondence between the electronic and superconducting linear response functions on each lattice site is maintained at all impurity concentration.

The low-temperature dependence of the superfluid density or penetration depth in both fresh (25 day-old) and aged (400 day-old) PuCoGa$_5$ are consistent with a line of nodes in a strongly disordered $d_{x^2-y^2}$-wave order parameter. The Swiss Cheese model can describe the quadratic temperature dependence of the superfluid density $\rho_{\rm s}(T)$ with the gap $\Delta(0)$ or $T_c$ reduced by about 20\% for impurity concentration $n_{\rm imp}=4\%$. It can account for at least a 70\% reduction in $\rho_{\rm s}(0)$ contrary to the dirty $d$-wave theory.
We attribute this to the fact that PuCoGa$_5$ possesses a relatively short coherence length, and, therefore, the conventional Abrikosov-Gor'kov pair-breaking theory, in which the order parameter is spatially averaged, is inappropriate. This result is similar to what is known for radiation damaged or doped high-$T_c$ cuprate superconductors.

For low impurity concentrations the short-coherence-length model agrees with the expected suppression of $\rho_{\rm s}(0)$ obtained from the dirty $d$-wave theory, while it deviates drastically otherwise.  Irrespective of the approach to impurity averaging, both theories predict a change from linear to quadratic in temperature in the superfluid density for large impurity concentrations. We show that the quadratic dependence of the superfluid density arises from the combined effects of the SC `gap filling' and `gap closing' in response to the presence of strong disorder.

Furthermore, both the fresh and aged PuCoGa$_5$ samples are consistent with the Swiss Cheese model for a weak-coupling gap $\Delta/T_c \sim 1.6-2.0$ [$\Delta=4\Delta(0)$, averaged over four NN sites for $d$-wave pairing] that is suppressed by strong impurity scattering. Finally, our lattice calculations show that although the order parameter is significantly suppressed in the immediate vicinity of impurities, and the superfluid density is strongly suppressed over extended regions along the nodal directions,  the superconductivity remains remarkably resilient. These calculations provide further evidence that PuCoGa$_5$ is the link between low-temperature heavy-fermion and high-temperature cuprate superconductors.

\section*{Acknowledgments}
We thank A. V. Balatsky and R. H. Heffner for discussions and encouraging this study.
Work at the Los Alamos National Laboratory was performed under the U.S.\ Department of Energy
contract No.\ DE-AC52-06NA25396 through the LDRD program and the Office of Basic Energy Sciences.
We used  computational resources of the National Energy Research Scientific Computing Center, which is supported by the Office of
Science of the U.S.\ DOE under Contract No.\ DE-AC02-05CH11231.


\begin{thebibliography}{10}
\bibitem{EinzelYEAR86} D. Einzel, P.J. Hirschfeld, F. Gross, B.S. Chandrasekhar, K. Andres, H.R. Ott, J. Beuers, Z. Fisk, and J.L. Smith, Phys. Rev. Lett. {\bf 56}, 2513 (1986).
%
\bibitem{GrossYEAR86} F. Gross, B.S. Chandrasekhar, D. Einzel, K. Andres, P.J. Hirschfeld, H.R. Ott, J. Beuers, Z. Fisk, and J.L. Smith, Z. Phys. B {\bf 64}, 175 (1986).
%
\bibitem{BarashYEAR96} Yu.S. Barash and A.A. Svidzinsky, Phys. Rev. B {\bf 53}, 15254 (1996).
%
\bibitem{AnnetBook} J. F. Annet, N. D. Goldenfeld, and S. R. Renn, {\it Physical Properties of High Temperature Superconductors II}, edited by D. M. Ginsberg (World Scientific, New Jersey, 1990).
%
%
\bibitem{ChoiYEAR88} C.H. Choi and P. Muzikar, Phys. Rev. B {\bf 37}, 5947 (1988).
%
\bibitem{KlemmYEAR88} R.A. Klemm, K. Scharnberg, D. Walker, and C.T. Rieck, Z. Phys. B {\bf 72}, 139 (1988).
%
\bibitem{AnnetPRB}  J. Annett, N. Goldenfeld, and S.R. Renn, Phys. Rev. B {\bf 43}, 2778 (1991).
%
\bibitem{ProhammerYEAR91} M. Prohammer and J. P. Carbotte, Phys. Rev. B
{\bf 43}, 5370 (1991).
\bibitem{ArbergYEAR93} P. Arberg, M. Mansor, and J. P. Carbotte, Solid
State Commun. {\bf 86}, 671  (1993).
%
\bibitem{HirschfeldYEAR93} P.J. Hirschfeld and N. Goldenfeld, Phys. Rev. B
{\bf 48}, 4219 (1993).
%
\bibitem{XuYEAR95} D. Xu, S.K. Yip, and J. A. Sauls, Phys. Rev. B {\bf 51}, 16233 (1995).
%
\bibitem{FranzYEAR97} M. Franz, C. Kallin, A. J. Berlinsky, and
M. I. Salkola, Phys. Rev. B {\bf 56}, 7882 (1997).
%
\bibitem{ByersYEAR93} J. M. Byers, M. E. Flatt\'{e}, and D. J. Scalapino, Phys. Rev. Lett. {\bf 71}, 3363 (1993).
%
\bibitem{BalatskyYEAR95} A. V. Balatsky, M. I. Salkola, and A. Rosengren, Phys. Rev. B {\bf 51}, 15547 (1995).
%
\bibitem{FlatteYEAR97} M. E. Flatt\'{e} and J. M. Byers, Phys. Rev. Lett. {\bf 78}, 3761 (1997).
%
\bibitem{SalkolaYEAR97} M. I. Salkola, A. V. Balatsky, and J. R. Schrieffer, Phys. Rev. B {\bf 55} 12648 (1997).
%
\bibitem{CSTing}Jian-Xin Zhu, T. K. Lee, C. S. Ting, and C.-R. Hu, Phys. Rev. B {\bf 61}, 8667 (2000).
%
\bibitem{BalatskyRMP} A. V. Balatsky, I. Vekhter, and Jian-Xin Zhu, Rev. Mod. Phys. {\bf 78}, 373 (2006).
%
\bibitem{UemuraYEAR91} Y. J. Uemura,  A. Keren, L. P. Le, G. M. Luke, B. J. Sternlieb, W. D. Wu, J. H. Brewer, R. L. Whetten, S. M. Huang, Sophia Lin, R. B. Kaner, F. Diederich, S. Donovan, G. Gr\"uner, and K. Holczer, Nature (London) {\bf 352}, 605 (1991).
%
\bibitem{NachumiYEAR96} B. Nachumi, A. Keren, K. Kojima, M. Larkin, G. M. Luke,
J. Merrin, O. Tchernysh{\oe}v, and Y. J Uemura, Phys. Rev. Lett. {\bf 77},
5421 (1996).
%
\bibitem{UemuraYEAR00} Y. J. Uemura, Physica C {\bf 341-348}, 2117 (2000).
%
\bibitem{BauerYEAR11} E. D. Bauer, Yi-feng Yang, C. Capan, R. R. Urbano, C. F. Miclea, H. Sakai, F. Ronning, M. J. Graf, A. V. Balatsky, R. Movshovich, A. D. Bianchi, A. P. Reyes, P. L. Kuhns, J. D. Thompson, and Z. Fisk, Proc. Natl. Acad. Sci. (USA) {\bf 108}, 6857 (2011).
%
\bibitem{JutierYEAR2005} F. Jutier, Ph.D. thesis, ITU Karlsruhe,
Germany, 2006, unpublished.
%
\bibitem{JutierYEAR05} F. Jutier, J.-C. Griveau, E. Colineau, J. Rebizant,
P. Boulet, F. Wastin, and E. Simoni, Physica B {\bf 359-361}, 1078 (2005).
%
\bibitem{JutierYEAR08} F. Jutier, G. A. Ummarino, J.-C. Griveneau, F. Wastin, E. Colineau, J. Rebizant, N. Magnani, and R. Caciuffo, Phys. Rev. B {\bf 77}, 024521 (2008).
%
\bibitem{CurroYEAR05} N. J. Curro, T. Caldwell, E. D. Bauer, L. A. Morales,
M. J. Graf, Y. Bang, A. V. Balatsky, J. D. Thompson, and
J. L. Sarrao, Nature {\bf 434}, 622 (2005).
%
\bibitem{OhishiYEAR07} K. Ohishi, R. H. Heffner, G. D. Morris, E. D. Bauer, M. J. Graf, J.-X. Zhu,
L. A. Morales, J. L. Sarrao, M. J. Fluss, D. E. MacLaughlin, L. Shu, W. Higemoto, and T. U. Ito,
Phys. Rev. B {\bf 76}, 064504 (2007).
%
\bibitem{MorrisYEAR06} G. D. Morris, R. H. Heffner, E. D. Bauer, L. A. Morales,
J. L. Sarrao, M. J. Fluss, D. E. MacLaughlin, L. Shu, and J. E. Anderson,
Physica B {\bf 374-375}, 180 (2006).
%
\bibitem{ScalapinoYEAR92} D. J. Scalapino, S. R. White, and S. C. Zhang,
Phys. Rev. Lett. {\bf 68}, 2830 (1992).
%
\bibitem{XiangYEAR95} T. Xiang and J. M. Wheatley, Phys. Rev. B{\bf 51},
11721 (1995).
%
\bibitem{TrivediYEAR96}  N. Trivedi, R. Scalettar, and M. Randeria,  Phys. Rev. B
{\bf 54}, R3756 (1996).
%
\bibitem{GhosalYEAR98} A. Ghosal, M. Randeria, and N. Trivedi, Phys. Rev. Lett.
{\bf 81}, 3940 (1998).
%
\bibitem{GhosalYEAR01} A. Ghosal, M. Randeria, and N. Trivedi, Phys. Rev. B {\bf 65}, 014501 (2001).
%
\bibitem{HurtYEAR05} D. Hurt, E. Odabashian, W. E. Pickett, R. T. Scalettar, F. Monaini, T. Paiva, and R. R. dos Santos,
Phys. Rev. B {\bf 72}, 144513 (2005).
%
\bibitem{SarraoYEAR02} J. L. Sarrao, L. A. Morales, J. D. Thompson,
B. L. Scott, G. R. Stewart, F. Wastin, J. Rebizant, P. Boulet,
E. Colineau, and G.H. Lander, Nature {\bf 420}, 297 (2002).
%
\bibitem{HiessYEAR07} A. Hiess, A. Stunault, E. Colineau, J. Rebizant, F. Wastin, N. Kernavanois, G. J. McIntyre,
M. T. Fernandez-Diaz, E. Leli\'{e}vre-Berna, J. A. Paix\~ao, and G. H. Lander,
J. Magnet. Magnet. Mater. {\bf 310}, 709 (2007).
%
\bibitem{HiessYEAR08} A. Hiess, A. Stunault, E. Colineau, J. Rebizant, F. Wastin, R. Caciuffo, and G. H. Lander, Phys. Rev. Lett. 100, 076403 (2008).
%
\bibitem{HottaYEAR03} T. Hotta and K. Ueda, Phys. Rev. B {\bf 67},
104518 (2003); T. Maehira, T. Hotta, K. Ueda, and A. Hasegawa,
Phys. Rev. Lett. {\bf 90}, 207007 (2003).
\bibitem{OpahleYEAR03} I. Opahle and P. M. Oppeneer, Phys. Rev. Lett. {\bf 90}, 157001 (2003).
%
\bibitem{OpahleYEAR04} I. Opahle, S. Elgazzar, K. Koepernik, and P. M. Oppeneer, Phys, Rev. B {\bf 70}, 104504 (2004).
%
\bibitem{TanakaYEAR04} K. Tanaka, H. Ikeda and K. Yamada,
J. Phys. Soc. Jpn. {\bf 73}, 1285 (2004).
%
\bibitem{ShickYEAR05} A. B. Shick, V. Jani\v{s}, and P. M. Oppeneer, Phys. Rev. Lett. {\bf 94}, 016401 (2005).
%
\bibitem{HottaYEAR06} T. Hotta and K. Ueda,   New J. Phys. {\bf 8}, 24 (2006).
%
\bibitem{OppeneerYEAR07} P. M. Oppeneer, A. B. Shick, J. Rusz, S. Leb\'{e}gue, and O. Eriksson, J. Alloys and Compounds {\bf 444-445}, 109 (2007).
%
\bibitem{ShickYEAR11} A. B. Shick, J. Rusz, J. Koloren\v{c}, P. M. Oppeneer, and L. Havela, Phys. Rev. B {\bf 83}, 155105 (2011).
%
\bibitem{BdGbook}P. G. de Gennes {\it Superconductivity of Metals and Alloys}, Westview Press, Reading, MA (1999).
%
\bibitem{Hirschfeld} D.V. Efremov, M.M. Korshunov, O.V. Dolgov, A.A. Golubov, and P.J. Hirschfeld, arXiv:1104.3840.
%
\bibitem{JianXin}Jian-Xin Zhu, Rong Yu, A. V. Balatsky, and Qimiao Si, arXiv:1103.3509.
%
\bibitem{DurstYEAR00} A.C. Durst and P.A. Lee, Phys. Rev B {\bf 62}, 1270 (2000).
%
\bibitem{Das_rho} Tanmoy Das, R. S. Markiewicz, and A. Bansil, Phys. Rev. Lett. {\bf 98}, 197004 (2007); Tanmoy Das, R. S. Markiewicz, and A. Bansil, J. Phys. Chem. Solids, {\bf 69}, 2963 (2008 ).
%
\bibitem{rho_quad} P.J. Hirschfeld and N. Goldenfeld, Phys. Rev B {\bf 48}, 4219 (1993); T. Xiang, C. Panagopoulos, and J. R. Cooper, Inter. J. Mod. Phys. B {\bf 12}, 1007 (1998).
%
\bibitem{YipYEAR92} S.-K. Yip and J. A. Sauls, Phys. Rev. Lett. (1992).
\bibitem{WalterYEAR99} H. Walter, W. Prusseit, R. Semerad, H. Kinder, W. Assmann, H. Huber, H. Burkhardt, D. Rainer, and J. A. Sauls,
Phys. Rev. Lett. {\bf 80}, 3598 (1999)
%
\bibitem{Ulm} Eric R. Ulm, Jin-Tae Kim, and Thomas R. Lemberger, Steve R. Foltyn, and Xindi Wu, Phys. Rev. B {\bf 51}, 9193 (1995).
%
\bibitem{Basov}  D. N. Basov, A. V. Puchkov, R. A. Hughes, T. Strach, J. Preston, T. Timusk, D. A. Bonn, R. Liang, and W. N. Hardy, Phys. Rev. B {\bf 49}, 12165 (1994).
%
\bibitem{Bernhard}C. Bernhard, J. L. Tallon, C. Bucci, R. De Renzi, G. Guidi, G. V. M. Williams, and Ch. Niedermayer, Phys. Rev. Lett. {\bf 77}, 2304 (1996).
%
\bibitem{Moffat} S. H. Moffat, R. A. Hughes, and J. S. Preston, Phys. Rev. B {\bf 55}, R14741 (1997).
%
\bibitem{PreostiYEAR94}Gianfranco Preosti, Heesang Kim, and Paul Muzikar, Phys. Rev. B {\bf 50}, 1259 (1994).
%
\bibitem{MoffatYEAR97} S. H. Moffat, R. A. Hughes, and J. S. Preston,
Phys. Rev. B {\bf 55}, 14741 (1997).
%
\bibitem{ZhitomirskyYEAR98} M. E. Zhitomirsky and M. B. Walker,
Phys. Rev. Lett. {\bf 80}, 5413 (1998).
%
\bibitem{HettlerYEAR99} M. H. Hettler and P. J. Hirschfeld, Phys. Rev. B
{\bf 59}, 9606 (1999).
%
\bibitem{BauerYEAR06} E. D. Bauer, F. Ronning, C. Capan, M. J. Graf,
D. Vandervelde, H. Q. Yuan, M. B. Salamon, D. J. Mixson,
N. O. Moreno, S. R. Brown, J. D. Thompson, R. Movshovich,
M. F. Hundley, J. L. Sarrao, P. G. Pagliuso, and
S. M. Kauzlarich, Phys. Rev. B {\bf 73}, 245109 (2006).
%
\end{thebibliography}
\end{document}